\documentclass[namedreferences,hyperref,optionalrh,solaromanenum]{spr-sola} 

\usepackage{graphicx}                    
\usepackage{color}                       

\newcommand{\aap}{{\it Astron. Astrophys.}}

\chardef\us=`\_

\begin{document}

\begin{frontmatter}

\title{Heavy Elements Abundances Inferred from the First Adiabatic Exponent in the Solar Envelope}

%
\author[addressref={1}]{\inits{V.A.}\fnm{Vladimir A. }\snm{Baturin}\orcid{0009-0000-7642-6786}}
\author[addressref={1},corref,email={avo@sai.msu.ru}]{\inits{A.V.}\fnm{Anna V. }\snm{Oreshina}\orcid{0009-0004-7969-1840}}
\author[addressref={2}]{\inits{G.}\fnm{Ga\"el }\snm{Buldgen}\orcid{0000-0001-6357-1992}}
\author[addressref={1}]{\inits{S.V. }\fnm{Sergey V. }\snm{Ayukov}\orcid{0009-0001-8859-3570}}
\author[addressref={3}]{\inits{V.K. }\fnm{Victor K. }\snm{Gryaznov}\orcid{0000-0003-4167-5090}}
\author[addressref={4,5}]{\inits{I.L. }\fnm{Igor L. }\snm{Iosilevskiy}}
\author[addressref={2}]{\inits{A. }\fnm{Arlette }\snm{Noels}}
\author[addressref={2}]{\inits{R. }\fnm{Richard }\snm{Scuflaire}}

%
\runningauthor{V.A. Baturin et al.}
\runningtitle{Heavy elements abundances}

\address[id={1}]{Sternberg Astronomical Institute, M.V. Lomonosov Moscow State University,  119234, Moscow, Russia}

\address[id={2}]{STAR Institute, Université de Liège, Liège, Belgium}
\address[id={3}]{Federal Research Center of Problems of Chemical Physics and Medicinal Chemistry RAS, Chernogolovka, Russia}
\address[id={4}]{Joint Institute for High Temperatures RAS, Moscow, Russia}
\address[id={5}]{Moscow Institute of Physics and Technology, Dolgoprudnyi, Russia}

\begin{abstract}
	
The first adiabatic exponent profile, noted ${{\Gamma }_{1}}$, computed along adiabatic coordinates ($T$, $\rho$) is in the focus of our study. Under conditions of almost fully ionized hydrogen and helium, the ${{\Gamma }_{1}}$ profile is quite sensitive to heavy elements ionization. ${{\Gamma }_{1}}$ decreases in regions where an element is partially ionized. The recent helioseismic structural inversion is obtained with an accuracy better than ${{10}^{-4}}$ in the most of the adiabatic convective zone that allows to study ionization variations. The aim is to determine the major heavy elements content in the solar convective zone. 
The method of our research is synthesis of the $\Gamma_1$ profile which is based on a linear combination of the contributions of individual heavy elements. The idea of the approach was proposed and justified by \citeauthor{Baturin_2022} (\aap \, \textbf{660}, A125, \citeyear{Baturin_2022}). 
We find the best approximation of the inverted profile ${{\Gamma }_{1}}$ adjusting the abundances of major elements (C, N, O, Ne), meanwhile the abundances of elements heavier than neon are fixed. We synthesize the theoretical ${{\Gamma }_{1}}$ profile using the SAHA-S equation of state, and are able to reproduce the inverted profiles with an accuracy of $(1-2)\cdot {{10}^{-5}}$. Total mass fraction of heavy elements found by this method is $Z=0.0148\pm 0.0004$. The oxygen logarithmic abundance is $8.70\pm 0.03$, carbon $8.44\pm 0.04$, nitrogen $8.12\pm 0.08$, and neon $8.17\pm 0.09$. The obtained estimations of oxygen and carbon agree with spectroscopic abundances by \citeauthor{Asplund_2009} (\aap \, \textbf{653}, A141, \citeyear{Asplund_2021}).
	
\end{abstract}

%
\keywords{Convection Zone; Helioseismology, Inverse Modeling;  Plasma Physics }

\end{frontmatter}

%
\section{Introduction}

The Sun is an unique laboratory for studying the properties of plasma under conditions that are difficult to achieve on Earth. This allows us to test and refine our understanding of plasma physics, as well as the chemical composition of our nearest star. The importance of such studies can hardly be overestimated both from the point of view of physics and astronomy, since the Sun is a kind of reference point for interpreting observations of other stars. 

The abundance of most elements in the Sun is determined by spectroscopic analysis of photospheric radiation. Exceptions are elements with high ionization potential, noble gases such as helium and neon, which cannot be determined from photospheric spectroscopy. The neon abundance is determined in the transition region and the solar corona. This value differs from the photospheric one due to the first ionization potential (FIP) effect, which, however, is not well-known. Therefore, to estimate the neon abundance in the photosphere, the Ne/O ratio in the corona and the oxygen abundance in the photosphere are used, assuming the Ne/O ratio to be the same in the corona and photosphere (see, for example, \citealp{Asplund_2021}).
In addition, important elements of the C, N, O group, which have a moderate but significant ionization potential, have a small number of spectral lines and are therefore determined with a noticeable error. The small number of lines makes their determination dependent on the atmospheric model, the accuracy of atomic parameters and other external assumptions. 

Therefore, the helioseismic approach represents an important additional source of information. An example of such an alternative method is the determination of helium content in the Sun. The first reliable helioseismic estimations of the helium content were obtained in the early nineties, (see e.g. \citealp{Vorontsov_1991, Christensen_Dalsgaard_1991}). Nowadays the commonly accepted helium abundance in the convective zone is based entirely on helioseismic analysis methods \citep{Basu_Antia_2004}. There were also heloseismic estimations of the total mass fraction of heavy elements, i.e. elements heavier than He \citep{Basu_Antia_2004, Vorontsov_2013, Buldgen_2017,  Buldgen_2024}.

Our work is also based on the helioseismic inversion of the profile of the first adiabatic exponent ${{\Gamma }_{1}}\left( r \right)$, i.e. its variations with depth in the convective zone. We determine not only the total mass fraction of heavy elements, but also the content of individual elements: oxygen, carbon, nitrogen, and neon. This became possible due to the fact that modern helioseismology methods allow one to obtain ${{\Gamma }_{1}}\left( r \right)$ with high accuracy, about ${{10}^{-4}}$. We compare these data with theoretical profiles calculated for various chemical compositions of the plasma, and select a composition at which the theoretical and helioseismic profiles agree best.

We use data from \cite{Buldgen_2024}, namely a helioseismic inversion based on model A2. The equation of state of model A2 is SAHA-S with a mixture of heavy elements from \citep{Asplund_2009}.  Model A2 is a ``starting model'': it goes through the iterative inversion of Ledoux discriminant and becomes a seismic model, that is its structure is corrected to fit the inversion results of density profile in the convective zone.  On the final step, the ${{\Gamma }_{1}}(r)$ is inverted on pressure $P(r)$ and density $\rho (r)$ sequence of points to get the full seismic thermodynamic coordinates. The inversions at each step are computed using the SOLA method from \citep{Pijpers_Thompson_1994}, adapted in the InversionKit software \citep{Reese_2016}. 

Our consideration is performed under the assumption that the chemical composition is constant inside the convective zone because of intensive mixing (see for example, \citealp{Cox_Giuli_1968}). We assume also that entropy is constant in the most part of the convective zone (see Figure~1 in \citealp{Baturin_2022} for illustration). 

\section{Method}

We analyze the chemical composition of the plasma in the convective zone of the Sun, based on the profile of the first adiabatic exponent ${{\Gamma }_{1}}\left( r \right)$. In a fully ionized plasma, ${{\Gamma }_{1}}=5/3$. If the plasma is partially ionized, then ${{\Gamma }_{1}}$ decreases in the regions of ionization of elements, and the decrease is proportional to the mass fraction of the ionized element. Thus, variations of the adiabatic exponent with radius can help to determine the chemical composition of the solar plasma. 

We will consider physical conditions characteristic for the lower part of the convective zone of the Sun, with radius coordinates $r=(0.73-0.93){{R}_{\odot }}$. This interval corresponds to temperatures  $\log T=5.60-6.31$. The ionization of K-electrons of oxygen, carbon, nitrogen, and neon occurs under these conditions \citep{Baturin_2022}.

\subsection{Theoretical Variety of $\Gamma_1$  Profiles on Adiabatic Curves}
\label{Sect_TheorVarG1}

Let us define one important object of our research. This is the ${{\Gamma }_{1}}$  profile on the adiabatic curve in thermodynamic variables $\left\{ T,\rho ,P \right\}$. Let us consider the equation of state for a fixed chemical composition. An adiabatic curve, or simply “adiabat” hereafter, is a one-dimensional curve $\mathcal{A}_S$ along which specific entropy $S$  is constant. The union of all adiabats coincides with the two-dimensional manifold of equilibrium states or with the surface of the thermal equation of state $P\left(T,\rho\right)$. A well-known geometric property of adiabats is that through each equilibrium point there passes one and only one adiabat. In other words, adiabats differing in entropy do not cross each other.

Projection of the adiabat on the plane $\left\{ P,\rho  \right\}$  gives the tangent ${{\Gamma }_{1}}={{\left. {\partial \log P}/{\partial \log \rho }\; \right|}_{S}}$ for any point of adiabat ${\mathcal{A}_{S}}$. Thus, we obtain a correctly and completely defined profile ${{\Gamma }_{1}}{{\left. \left( T,\rho  \right) \right|}_{S}}$, with the parameter of specific entropy $S$. Noticeable that the primary element of this definition is the adiabatic curve ${\mathcal{A}_{S}}$.
The set of all theoretical ${{\Gamma }_{1}}$ profiles for different chemical composition are equivalent to the thermal and caloric equations of state.

In its most general form, the task of the study is to find a theoretical profile ${{\Gamma }_{1}}\left( {\mathcal{A}_{S}} \right)$ that will be closest to the profile obtained in the inversion procedure.

Although the chemical composition is determined specifically for a leaf of adiabats, the union of adiabats for all chemical compositions is a much more complicated geometrical complex. In practice we move to a rough concept, namely ${{\Gamma }_{1}}$ profile calculated on a fixed set of points $\left\{ T,\rho  \right\}$. Moreover, such a set of points just only approximately reproduces its own adiabat and may not be consistent with ${{\Gamma }_{1}}$ as a projection of the adiabat. The important thing in the LS-algorithm that is all calculations for different chemical compositions were carried out at the same points $\left\{ T,\rho  \right\}$ .

\subsection{Theoretical ${{\Gamma }_{1}}$ Profile in Solar Models}

We start by illustrating how mass fractions of individual elements affect the ${{\Gamma }_{1}}$ profile. We consider the mixtures given in the works by \cite{Grevesse_Noels_1993, Asplund_2009, Magg_2022}. The calculations of equation of state (EOS) are limited to the eight most abundant heavy elements: C, N, O, Ne, Mg, Si, S, Fe. We designate them GN93, AGSS09, MB22 to be short (see Appendix~\ref{Appendix_StandardMixtures}).

Figure 1 shows the profiles of the adiabatic exponent ${{\Gamma }_{1}}\left( r \right)$ in standard solar models, calculated with different contents of heavy elements (solid curves). The larger total mass fraction $Z$ of heavy elements in a model, the smaller ${{\Gamma }_{1}}$. In addition, ${{\Gamma }_{1}}\left( r \right)$ depends on relative mass fractions of individual elements in the mixture. The band between the solid curves specifies the possible range of ${{\Gamma }_{1}}$ profiles in solar models. 

The dashed black curve shows the ${{\Gamma }_{1}}$ profile for hydrogen-helium plasma without heavy elements. The profile is practically the same in different models, since $Z=0$. The minor difference between the sets of points $\left( T(r),\rho (r) \right)$ from the different models does not affect the $\Gamma _{1}^{\mathrm{HHe}}$  profile on this scale of the figure.

\begin{figure} 
\centerline{\includegraphics[width=1.0\textwidth,clip=]{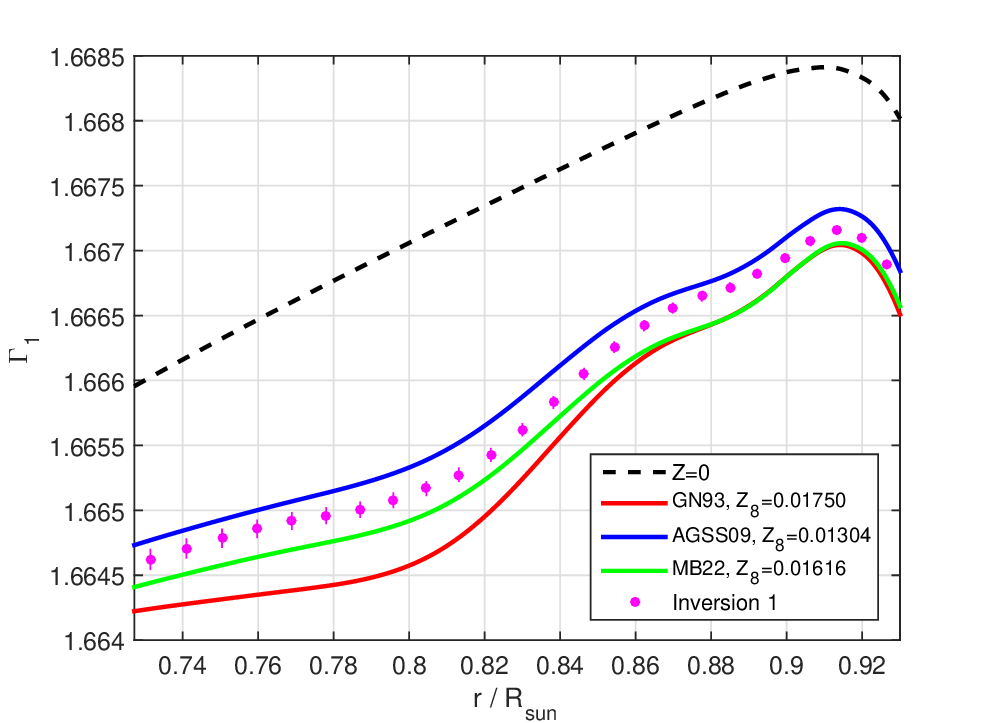}}
\caption{Adiabatic exponent in standard solar models with mixtures GN93, AGSS09, and MB22 (see Table~\ref{Table_mixtures} in Appendix~\ref{Appendix_StandardMixtures}). Black dashed curve is for hydrogen-helium plasma without heavy elements. Helium mass fraction in the convective zone $Y=0.2485$ in all models. The magenta dots show the results of helioseismic inversion \citep{Buldgen_2024}.}
\label{Fig_G1_mod_inv}
\end{figure}

\subsection{Definition of $Z$ Contribution and its Main Feature}

The object of study is not the adiabatic exponent ${{\Gamma }_{1}}$ itself, but the $Z$ contribution ${{\delta }_{Z}}{{\Gamma }_{1}}={{\Gamma }_{1}}-\Gamma _{1}^{\mathrm{HHe}}$, where $\Gamma _{1}^{\mathrm{HHe}}$ is the adiabatic exponent in a hydrogen-helium plasma. Thus, the $Z$ contribution is the decrease in ${{\Gamma }_{1}}$ due to the ionization of heavy elements and does not include the influence of hydrogen and helium, as well as Coulomb nonideality \citep{Baturin_2022}.

\begin{figure} 
	\centerline{\includegraphics[width=1.0\textwidth,clip=]{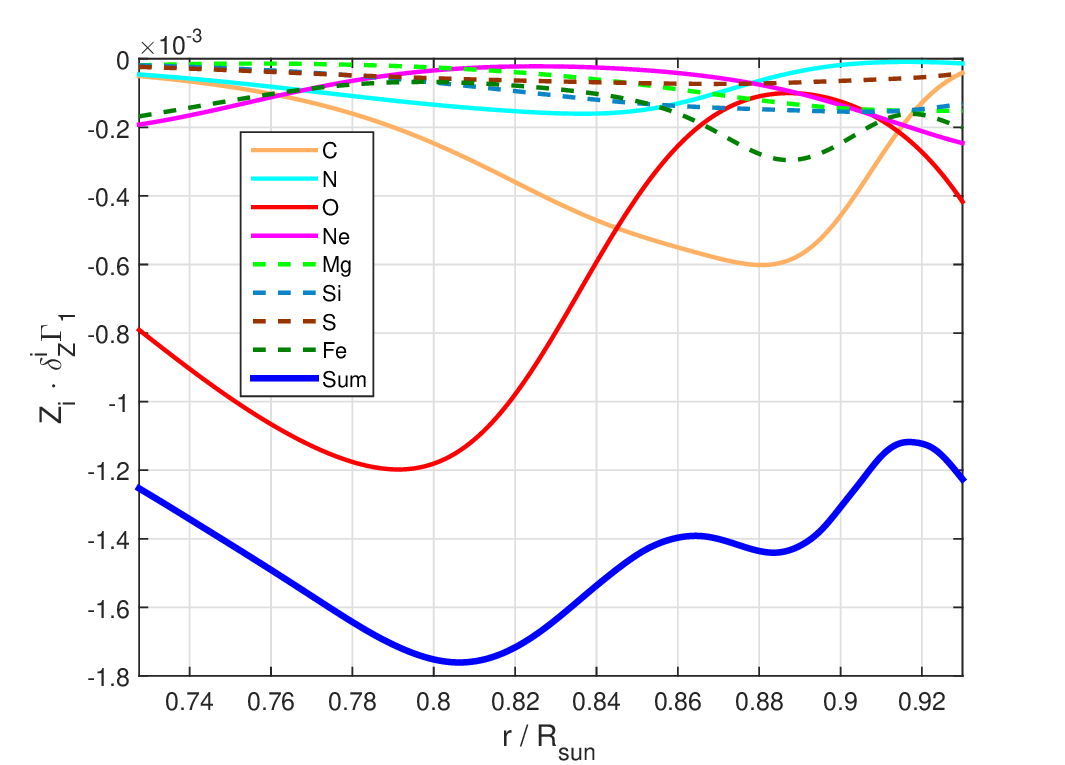}}
	\caption{$Z$ contribution in the mixture AGSS09 and contributions of individual elements.}
	\label{Fig_dG1_elements}
\end{figure}

Figure~\ref{Fig_dG1_elements} gives an example of the $Z$ contribution in the AGSS09 mixture (thick blue curve). In addition, this figure shows the contributions of individual elements according to their abundances in the AGSS09 mixture (see Table~\ref{Table_mixtures} in Appendix~\ref{Appendix_StandardMixtures}). \cite{Baturin_2022} showed that the total $Z$ contribution $\delta_Z\Gamma_1$ can be represented as the sum of the contributions of individual elements $\delta_Z^{(i)}\Gamma_1$, and each individual contribution is proportional to the mass fraction ${{Z}_{i}}$ of the element in the mixture:
\begin{equation}
\label{Eq_dZG1}
{{\delta }_{Z}}{{\Gamma }_{1}}=\sum \limits_{i=1}^{8}{{{Z}_{i}}\cdot \delta _{Z}^{(i)}{{\Gamma }_{1}}}.
\end{equation}

\noindent This is a basic equation of our approach. Here basis $\delta _{Z}^{(i)}{{\Gamma }_{1}}$ was calculated within the framework of the SAHA-S equation of state (\citeauthor{Gryaznov_2004} \citeyear{Gryaznov_2004}, \citeyear{Gryaznov_2006}, \citeyear{Gryaznov_2013}; see also the SAHA-S website \url{crydee.sai.msu.ru/SAHA-S_EOS}); the calculations were performed for a mixture in which one percent by mass is represented by only one heavy element (oxygen or carbon, or nitrogen, etc.), and the remaining part is hydrogen and helium.
The $Z$ contribution of an individual element shows specific minimum. This is due to the fact that the ionization of K electrons is isolated from the ionization of all other electrons. The temperature, at which K-ionization occurs, is specific for each of the elements of the C, N, O, Ne group and they appear inside the studied interval \citep{Baturin_2022}. The amplitude of the minimum is proportional to the content of the element. Oxygen makes the largest contribution, since it is the most abundant in solar plasma after hydrogen and helium. The next most important are carbon, nitrogen and neon. As can be seen from the figure, the minimum of the total $Z$ contribution at $r=0.80$ is associated mainly with oxygen, and at $r=0.88$ with carbon.

\begin{figure} 
	{\includegraphics[width=1.0\textwidth,clip=]{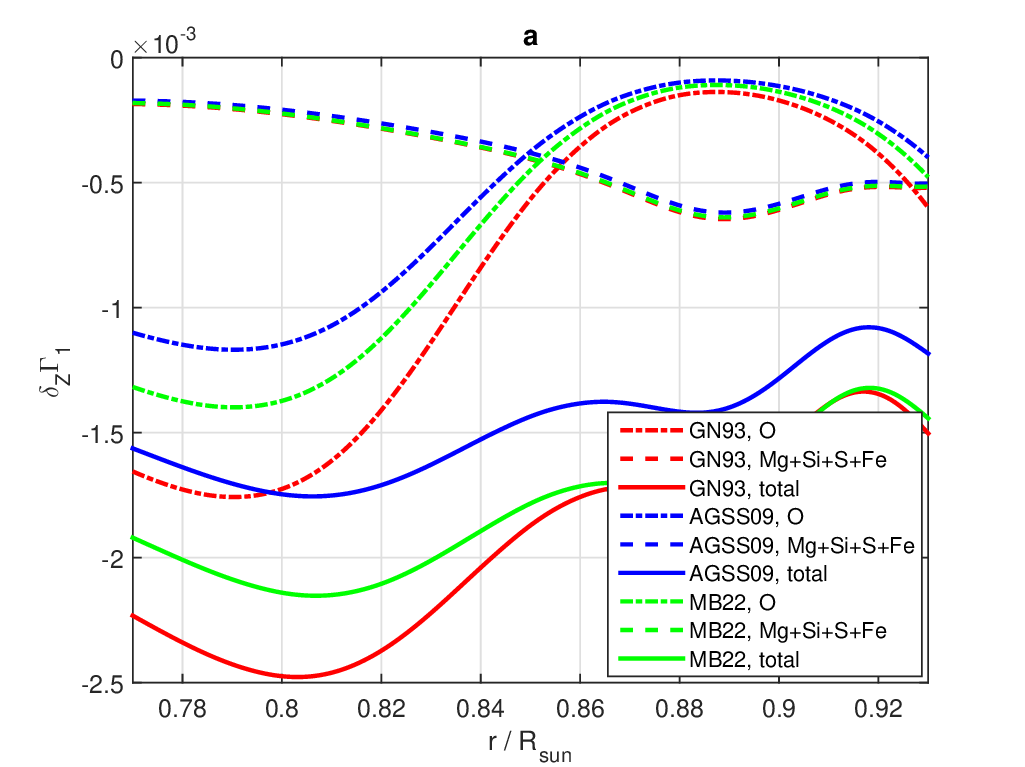}}
	{\includegraphics[width=1.0\textwidth,clip=]{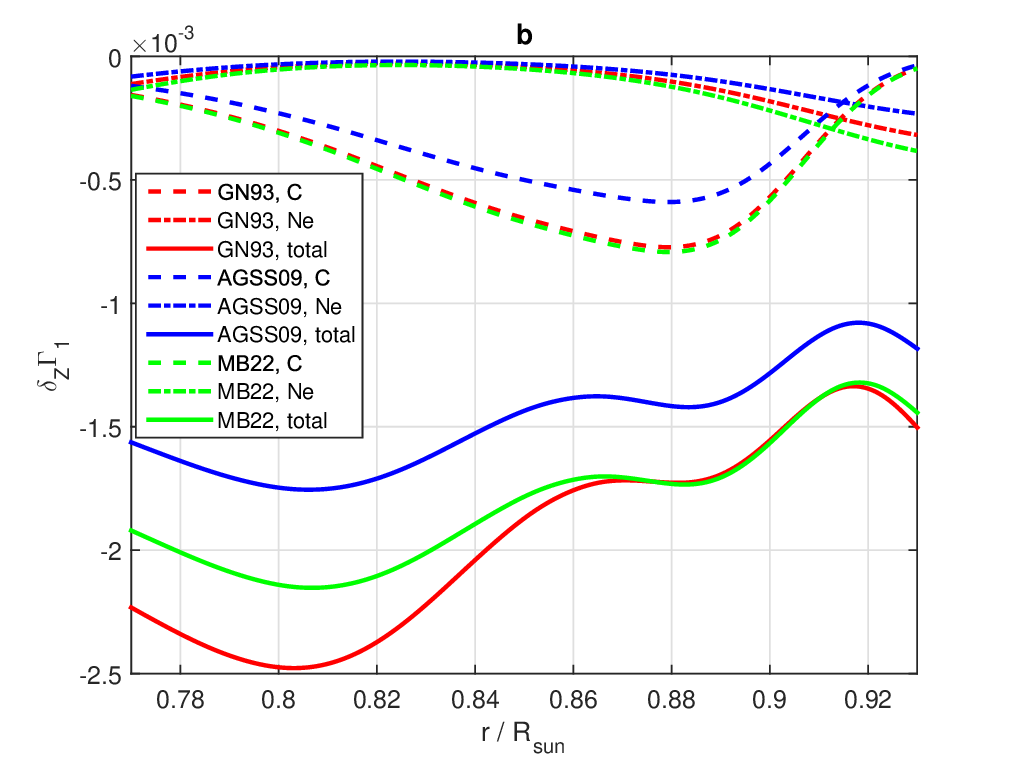}}
	\caption{$Z$ contributions of elements and total $Z$ contributions (solid curves) in mixtures GN93, AGSS09, MB22. (a) $Z$ contributions of oxygen (dash-dotted curves) and Mg+Si+S+Fe (dashed curves).  (b) $Z$ contributions of carbon (dashed curves), nitrogen (thin solid curves), neon (dash-dotted curves).}
	\label{Fig_dG1_elements_mixtures}
\end{figure}

$Z$ contributions in different mixtures are presented in Figure~\ref{Fig_dG1_elements_mixtures} by solid curves. The $Z$ contribution of four elements (Mg, Si, S, Fe) is almost the same in different mixtures (dashed curves in Figure~\ref{Fig_dG1_elements_mixtures}(a)), while the contributions of oxygen is noticeably different in the region $r=0.73-0.84$ (dash-dotted curves in Figure~\ref{Fig_dG1_elements_mixtures}(a)). 

Thus, we see that different mass fractions of elements in mixtures lead to different profiles of $Z$ contributions ${{\delta }_{Z}}{{\Gamma }_{1}}(r)$. And vice versa, by analyzing a profile ${{\delta }_{Z}}{{\Gamma }_{1}}(r)$, we can draw a conclusion about mass fractions of individual elements. This idea underlies our method for analyzing the chemical composition of the plasma in the solar convective zone.

The determination of $Z$ contributions and the study of their properties were performed by \cite{Baturin_2022} under the assumption of fixed temperature and density. This means that $\Gamma_1$ for helium and hydrogen, as well as $\Gamma_1$ for the contributions of individual elements, are calculated for the same $T$, $\rho$ within the equation of state. These are theoretical functions that have no relation to either the real Sun or its models. 

At the same time, the pressure value for each value of $\Gamma_1$ is different due to the fact that the chemical composition changes (H+He, H+He+C, H+He+O, …). Therefore, the choice of fixed variables in the form of $P$, $\rho$, generally speaking, will give a slightly different result for the contribution functions. We choose $T$, $\rho$ as a more stable algorithm.

\subsection{Algorithm}

The input data for our method is the profile of ${{\Gamma }_{1}}$ obtained using helioseismic inversion. The basis functions $\delta _{Z}^{i}{{\Gamma }_{1}}$ are assumed to be known functions of temperature $T$, density $\rho $, and hydrogen mass fraction $X$,  and are calculated using the SAHA-S equation of state. The output data are the mass fractions ${{Z}_{i}}$, which are estimated using the weighted least squares (LS)  method to best approximate the inversion profile. The algorithm is described in detail by \cite{Baturin_2022}. We fixed the mass fraction of four elements (Mg, Si, S, Fe) as they are given in AGSS09. This approach is justified by the fact that the mass fraction of these elements remains almost unchanged in the estimations of different authors, in contrast to oxygen, for example (see Figure~\ref{Fig_dG1_elements_mixtures}a). The LS synthesis of $Z$ contribution is performed using four basis functions (C, N, O, Ne). As a result, we obtain the mass fractions ${{Z}_{i}}$ of carbon, nitrogen, oxygen, and neon, as well as the total mass fraction $Z=\sum\limits_{i=1}^{8}{{{Z}_{i}}}$  of eight heavy elements (C, N, O, Ne, Mg, Si, S, Fe).  

We obtain this $Z$ estimation from the general decrease of ${{\Gamma }_{1}}$ on the Sun. Therefore, it is an estimation of the mass abundance of all heavy elements, and not just the eight selected ones. Mass fractions of the individual elements C, N, O, and Ne may be slightly overestimated, because they include mass fractions of discarded elements distributed over this group. However, this overestimation is insignificant, since the total contribution of all discarded elements does not exceed $4\times {{10}^{-4}}$  (see Appendix~\ref{Appendix_StandardMixtures}). 

The accuracy of the LS approximation is estimated by standard deviation $\varepsilon $, that is the difference between $Z$ contributions obtained by our LS method and from the inversion data:
\begin{equation}
\varepsilon =\sqrt{\frac{1}{N}\sum\limits_{j=1}^{N}\left[ {{\delta }_{Z}}{{\Gamma }_{1}}^{\mathrm{LS}}-{{\delta }_{Z}}{{\Gamma }_{1}}^{\mathrm{inv}} \right]^{2}}.
\label{Eq_epsilon} 
\end{equation}

\noindent Here $N$ is number of points in inversion profile ${{\Gamma }_{1}}\left( r \right)$. The smaller $\varepsilon $, the better the agreement between theory and observations.

\subsection{Approximation Model to Connect Thermodynamic and Seismic Data }
\label{SbSection_ApproxRho}

As described above, our method is based on the thermodynamic synthesis of the profile $\Gamma_1$ on a sequence of points $T$, $\rho$, which should approximate solar conditions. The key point is the calculation of the $Z$ contribution $\delta_Z\Gamma_1^{\mathrm{inv}}$ for the inverted profile $\Gamma_1^{\mathrm{inv}}$. For this, we must be able to calculate $\Gamma_1^{\mathrm{HHe}}$, which will be further subtracted from the inverted profile. The profile $\Gamma_1^{\mathrm{HHe}}$ is purely theoretical, based on our assumptions about the structure of the solar model, as well as on the equation of state. 

The profile $\Gamma_1^{\mathrm{HHe}}$ is calculated on a sequence of points $T$, $\rho$, with assumed hydrogen mass fraction $X$. On the other hand, seismic inversion is able to provide information about the structure of the model in the variables $P$ and $\rho$ \citep{Buldgen_2024}. To build a bridge between these two representations, we use the method of an approximation envelope. Based on the inversion data \citep{Buldgen_2024}, we calculate an envelope model which accurately reproduces the inverted density and pressure profiles. The key parameter of the envelope for adjusting the density profile is the mass coordinate at $r = 0.75R_{\odot}$ which is named $m_{75}$. An example of envelope fitting is given in Appendix~\ref{Appendix_Density_approximation}. The parameters of the approximation model are $m_{75} = 0.98236$, $X = 0.7377$, $Y = 0.2479$, $Z = 0.0144$. The accepted $m_{75}$ agrees with the value $m_{75} = 0.9822 \pm 0.0002$ found by \cite{Vorontsov_2013}. 

After constructing such a model, we obtain the required $T(r)$ profile, which is used to calculate both the $\Gamma_1^{\mathrm{HHe}}$ profile and the basic $Z$ contributions $\delta_Z\Gamma_1^{(i)}$ of individual elements. The most problematic parameter of the approximation model is the chemical composition, or more precisely, the hydrogen content $X$. We study the influence of X on our final results in Appendix~\ref{Appendix_Methodical_errors}.

It is worth noting that this kind of temperature data may be different when using any alternative theoretical EOS model.

\subsection{$Z$ Contribution Obtained from Inversion }

\begin{figure} 
	\centerline{\includegraphics[width=1.0\textwidth,clip=]{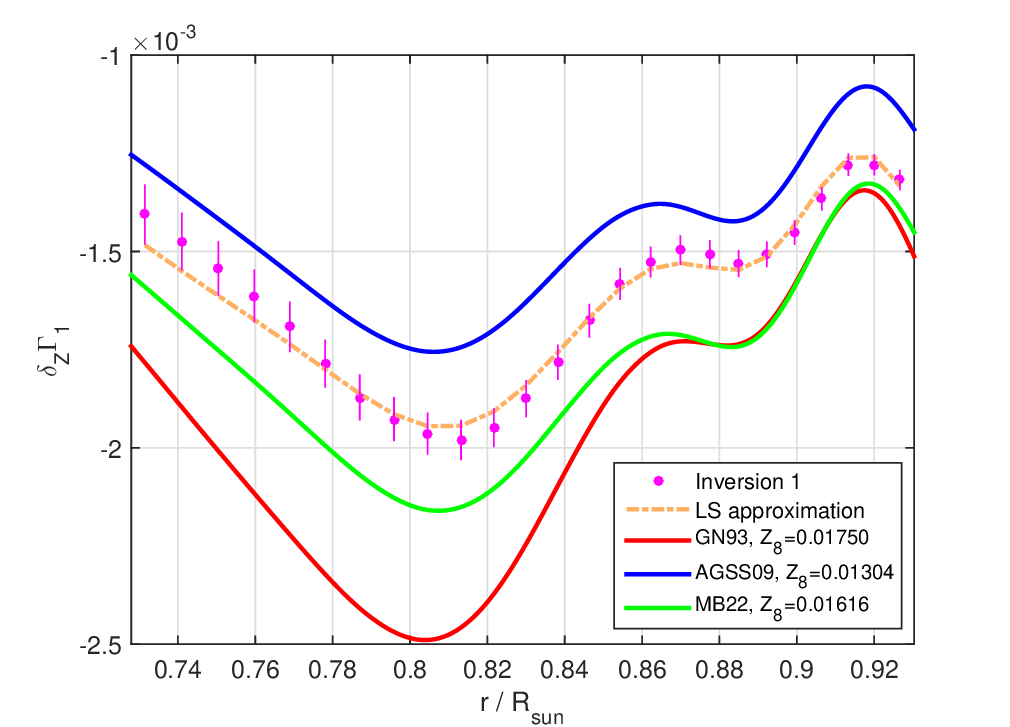}}
	\caption{$Z$ contribution in inversion (magenta points with invertion errors), in theoretical LS approximation (orange dot-dashed curve), and in solar models calculated with different mixtures (red, blue and green curves).}
	\label{Fig_dG1_inversion}
\end{figure}

The inversion profile ${{\Gamma }_{1}}\left( r \right)$ was obtained by \cite{Buldgen_2024} and used in our study. From the inverted ${{\Gamma }_{1}}\left( r \right)$ a theoretical background $\Gamma _{1}^{\mathrm{HHe}}\left( T\left( r \right),\rho \left( r \right) \right)$ was subtracted to obtain $Z$ contribution in the Sun.  Figure~\ref{Fig_dG1_inversion} shows the $Z$ contribution ${{\delta }_{Z}}{{\Gamma }_{1}}^{\mathrm{inv}}$ by magenta dots. Vertical bars demonstrate inversion errors given by \cite{Buldgen_2024} which came from observational errors. 

$Z$ contributions in models with standard mixtures are shown by solid lines for comparison with ${{\delta }_{Z}}{{\Gamma }_{1}}^{\mathrm{inv}}$. The $Z$ contribution of inversion ${{\delta }_{Z}}{{\Gamma }_{1}}^{\mathrm{inv}}$ is located between curves for the models computed with mixtures AGSS09 and MB22. It is reasonable to expect that the total $Z$ content in the Sun also is between the $Z$ values for these two mixtures. In addition, the shape of the inverted $Z$ contribution is close to the shape of the theoretical curves. This allows us to hope that such an inverted $Z$ contribution can be adequately approximated with some theoretical curves for specifically adjusted mixture. If they would be not similar, the approximation could not work. 

\section{Results}

Using the inverted ${{\delta }_{Z}}{{\Gamma }_{1}}^{\mathrm{inv}}$, we found the theoretical profile ${{\Gamma }_{1}}$  and the corresponding $Z$ contribution using the weighted LS  method so that the approximation error is minimal. The orange dot-dashed curve in Figure~\ref{Fig_dG1_inversion} shows the result ${{\delta }_{Z}}{{\Gamma }_{1}}^{\mathrm{LS}}$ of the LS approximation. The approximation is within the inversion errors (about ${{10}^{-4}}$) and the standard deviation $\varepsilon =3.6\cdot {{10}^{-5}}$ (Equation~\ref{Eq_epsilon}).

The obtained approximation ${{\delta }_{Z}}{{\Gamma }_{1}}^{\mathrm{LS}}$ gives elements abundances in the adiabat space in the theoretical equation of state, which is best compared with the inverted profile ${{\Gamma }_{1}}^{\mathrm{inv}}$. Thus the main task of the study is performed. 

The approximated ${{\delta }_{Z}}{{\Gamma }_{1}}^{\mathrm{LS}}$ gives the total mass fraction of heavy elements. Indeed, the approximation has the form of Equation~\ref{Eq_dZG1} from which we can get  $Z=\sum\limits_{i}^{{}}{{{Z}_{i}}}$. We estimate $Z=0.0148\pm 0.0004$. The error of this value was estimated from the inversion error of ${{\Gamma }_{1}}$, as well as from the variation of model parameters (see Appendix~\ref{Appendix_Methodical_errors}). Our result lies between the estimates in \citep{Asplund_2009} and \citep{Magg_2022} that agrees with qualified analysis of $Z$ contribution in Figure~\ref{Fig_dG1_inversion}. This result also coincides with interval estimation $Z=0.0131-0.0151$ for Data Set 1 by \cite{Buldgen_2024}.

In addition, the theoretical approximation allows us to determine the mass fractions of individual elements. The results are presented in Table~\ref{Table_Main_result}.

\begin{table}
\caption{Mass fractions ${{Z}_{i}}$ of elements, as well as logarithmic abundances $\log {{A}_{i}}$, obtained as a result of weighted least squares approximation of inverted $Z$ contribution.}
\label{Table_Main_result}
\begin{tabular}{lcc}     
\hline
Element &	${{Z}_{i}}$ &$\log {{A}_{i}}$ \\
\hline
C  & 0.00244 & 8.444 \\
N  & 0.00135 & 8.121 \\
O  & 0.00590 & 8.702 \\
Ne & 0.00217 & 8.166 \\
\hline
$Z$ & 	\multicolumn{2}{c}{0.01485} \\
$\varepsilon$ & \multicolumn{2}{c}{$3.6\cdot {{10}^{-5}}$ } \\
\hline 
\end{tabular}
\end{table}

Oxygen and carbon are determined most confidently. We estimate the oxygen mass fraction ${{Z}_{\mathrm{O}}}=0.0059$ and carbon mass fraction ${{Z}_{\mathrm{C}}}=0.0024$. Relative errors $\Delta {{Z}_{i}}/{{Z}_{i}}$ does not exceed 5\% (see Appendix~\ref{Appendix_Methodical_errors}). 

Spectroscopy uses logarithmic abundances $\log A$ that is the relative number $N$  of atoms of some element $E$  compared to the number of hydrogen atoms $N(H)$  in a logarithmic scale:

\begin{equation}
\log {{A}_{E}}={{\log }_{10}}\frac{N(E)}{N(H)}+12.
\end{equation}

\noindent The oxygen logarithmic abundance is $8.70\pm 0.03$, carbon $8.44\pm 0.04$. 

The theoretical approximation also produces estimations of nitrogen and neon. They are not determined with a very high robustness, but they are an integral part of the LS approximation. Their mass fractions are ${{Z}_{\mathrm{N}}}=0.0013$ and ${{Z}_{\mathrm{Ne}}}=0.0022$. The relative errors $\Delta {{Z}_{i}}/{{Z}_{i}}$ achieve 16\% for nitrogen and 20\% for neon. Logarithmic abundances are for nitrogen $8.12\pm 0.08$, and neon $8.17\pm 0.09$.

\section{Comparison with Spectroscopic Results}

\begin{figure} 
\centerline{\includegraphics[width=1.0\textwidth,clip=]{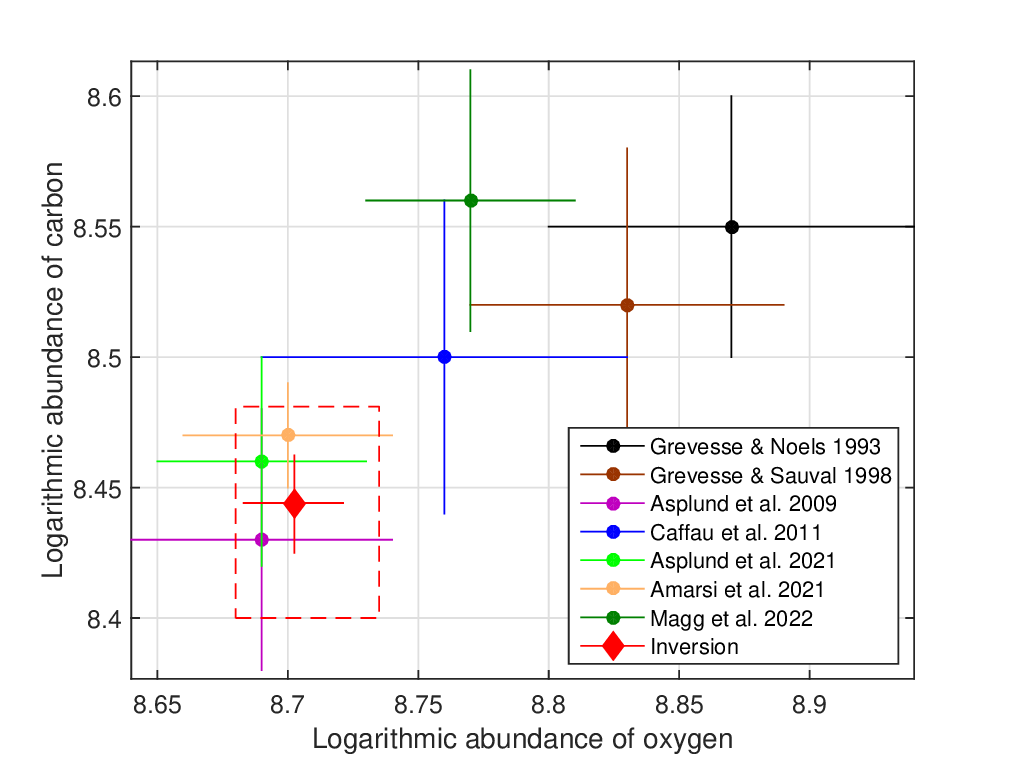}}
\caption{Logarithmic abundances of oxygen and carbon obtained using spectroscopy (circles) and from the helioseismic inversion (diamond). Crosses show errors of the estimations. Red dashed rectangle describes domain of possible errors of our estimation due to uncertainties of input parameters.}
\label{Fig_lgA_O_C}
\end{figure}

\begin{figure} 
	\centerline{\includegraphics[width=1.0\textwidth,clip=]{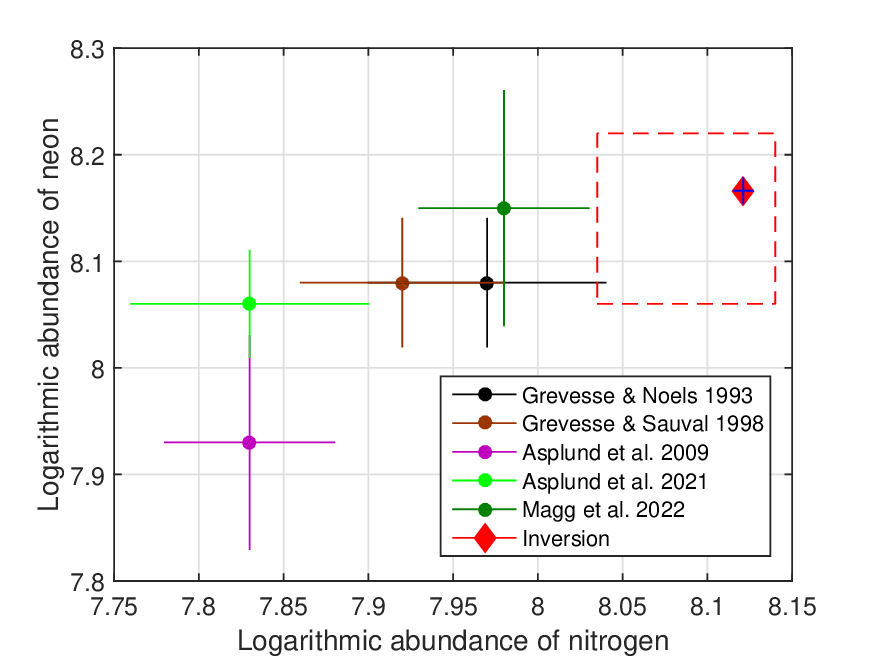}}
	\caption{Logarithmic abundances of nitrogen and neon obtained using spectroscopy (circles) and from the helioseismic inversion (diamond). Crosses show errors of the estimations. Red dashed rectangle describes domain of possible errors of our estimation due to uncertainties of input parameters.}
	\label{Fig_lgA_N_Ne}
\end{figure}

Spectroscopic estimations of the logarithmic abundance of oxygen, carbon, nitrogen and neon, given by different authors, are shown in Figures~\ref{Fig_lgA_O_C} and \ref{Fig_lgA_N_Ne} by circles. The crosses indicate error ranges. In addition, our estimation obtained from the helioseismic inversion is shown by red diamond. The cross next to the diamond shows the errors arising from the errors $\Delta {{\Gamma }_{1}}$, which are given by the inversion. The large red dashed rectangle outlines the possible range of errors that arise due to varying internal parameters of our method: temperature profile $T(r)$, hydrogen mass fraction $X$, the total contribution of Mg, Si, S and Fe. These errors are explored in more detail in Appendix~\ref{Appendix_Methodical_errors}.

Oxygen is determined most confidently in the ${{\Gamma }_{1}}$ analysis due to the isolation of the contribution of oxygen and its high abundance. The resulting logarithmic abundance is $\log {{A}_{\mathrm{O}}}=8.70$. This estimation is slightly higher than in \citeauthor{Asplund_2009} (\citeyear{Asplund_2009}, \citeyear{Asplund_2021}), and is in perfect agreement with the results by \cite{Amarsi_2021}. On the other hand, the agreement with
\cite{Caffau_2011} and \cite{Magg_2022} is really marginal. The obtained estimation falls within the range of errors of the last two works.  However, our estimation does not agree with the values by \cite{Grevesse_Noels_1993} and \cite{Grevesse_Sauval_1998}. The overall conclusion for oxygen is that we 
heavily favour \citeauthor{Asplund_2009} (\citeyear{Asplund_2009}, \citeyear{Asplund_2021}) and \cite{Amarsi_2021}.

Carbon is intermediate between the spectroscopic estimations by \citeauthor{Asplund_2009} (\citeyear{Asplund_2009}, \citeyear{Asplund_2021}), being statistically consistent with both results and falling within the error interval. Our estimation is $\log {{A}_{\mathrm{C}}}=8.44$, while AGSS09 gives $\log {{A}_{\mathrm{C}}}=8.43$. Other spectroscopic estimations of carbon are higher, although our result agrees within errors with the result by \cite{Caffau_2011}, \cite{Amarsi_2021}, and \cite{Grevesse_Sauval_1998}. However, we can confidently isolate ourselves from the estimates by \cite{Magg_2022} and \cite{Grevesse_Noels_1993}. 

Thus, our estimations of oxygen and carbon are in good agreement with \citeauthor{Asplund_2009} (\citeyear{Asplund_2009}, \citeyear{Asplund_2021})  and \cite{Amarsi_2021} and hint at slightly increased oxygen and decreased carbon.

Nitrogen is determined by our method not very reliably, which is explained by its low mass fraction. In any case, we get the nitrogen abundance, $\log {{A}_{\mathrm{N}}}=8.12$. This abundance is higher than any spectroscopic estimation. The highest spectroscopic result is by \cite{Magg_2022}, which is 7.98.

Neon, like nitrogen, is not determined very robustly. The contribution of neon to ${{\delta }_{Z}}{{\Gamma }_{1}}(r)$ appears only at the edges of the working interval (see magenta curve in Fig.~\ref{Fig_dG1_elements}); possible errors are quite noticeable. However, our result is consistent with the values given by \cite{Grevesse_Noels_1993, Grevesse_Sauval_1998, Asplund_2021, Magg_2022} and currently accepted ones. Our estimations yield $\log {{A}_{\mathrm{Ne}}}=8.17$, which is significantly higher than value $7.93\pm 0.10$ given by \cite{Asplund_2009}.

\section{Discussion}
Our study focuses on a method to extract information about the content in individual elements from the thermodynamic properties of the ionized solar plasma by sequential filtering the $\Gamma_1$ profile. The initial and a priori information for our method is the inversion profile $\Gamma_1$, whereas the accuracy and procedure of obtaining such a seismic profile are beyond the scope of our article. We do not consider here the whole problem of the reconstruction of a seismic model of the Sun that would match the observed frequencies. 
	
The proposed method consists of subtracting the $\Gamma_1^{\mathrm{HHe}}$ contribution from the total $\Gamma_1$, as well as fixing the $Z$ contribution of the group (Mg, Si, S, Fe), which allows us to obtain a stable estimate of the carbon and oxygen contents in the convective zone of the Sun. In Section~\ref{Sect_TheorVarG1}, a set of self-consistent adiabatic $\Gamma_1$ profiles was defined within the framework of the equation of state. The condition for the application of our method is the reasonable proximity of the analyzed inversion profile to one of the profiles of the theoretical set. In this case, the method provides information about the oxygen and carbon contents. If the inversion profile is very different from any theoretical ones, then our method cannot be applied.	
	
The main challenge of the obtained contents relies in the determination of the $\Gamma_{1}$ profile in an accurate and precise way. The approach relies on multiple successive structure inversions \citep[see][for details]{Buldgen_2024} which are not exempt from small inaccuracies. Improving the accuracy and reliability of the $\Gamma_{1}$ determination will directly impact the accuracy and reliability of our investigation on individual abundances. 
	
The first uncertainty results in the direct use of individual frequencies, that lead to the pollution of the inversion results by the so-called “surface effects”. This implies that the use of high-degree modes, which may help in better resolving the outer layers of the convective envelope \citep{Reiter_2020}, will require an adaptation of the surface correction employed in the inversion, using the degree dependent formulation of \cite{DiMauro2002} instead of the usual polynomial correction of \citet{Rabello_1999}. Tests using this correction should be carried out even with the current set of frequencies using degree as high as 250.

Another point that would mitigate the surface effects would be the direct use of averaged hydrodynamical simulations for the outer envelope, using a better depiction of the stratification of the upper convective layers and reduce naturally the surface effects. Another approach to mitigate this would be to use the entropy-calibrated values for the mixing-length parameter that has been implemented in some stellar evolution codes \citep{Manchon_2024}.

Last but not least, the effect of the radius uncertainty should be taken into account when carrying out the successive inversions for the structure \citep{Basu_1998}. This could be done by using the appropriate kernels provided in \cite{Takata_2024}, or additional terms as done in \citep{Richard_1998}. Given that our calibrated radius value matches theirs very precisely, we can estimate that the overall impact on the final results will be limited, but that some impact on the uncertainties may be expected due to larger uncertainties on the $\Gamma_{1}$ inversion. Further investigations are required to fully quantify the impact of all these effects on the $\Gamma_{1}$ inversion to see how far can these structural inversions be pushed in abundance determinations.

\section{Conclusion}

The method of theoretical approximation of the adiabatic exponent profile is applied to data obtained from helioseismic inversion. We show how relative variations in chemical composition can explain features in the helioseismically inverted ${{\Gamma }_{1}}$ profile. We assumed the mass fraction of elements Mg, Si, S, and Fe to be known and equal to those determined in spectroscopic observations. The main results are the following.

The inverted ${{\Gamma }_{1}}$ profile is reproduced by the theoretical one under accuracy $(2-4)\cdot {{10}^{-5}}$ with adjusting the fractions of four major elements (O, C, Ne, N).

The total mass fraction of heavy elements is $Z=0.0148$. Our result lies between the estimates  $0.0134$ by \cite{Asplund_2009} and $0.0165$ by \cite{Magg_2022} and coincides with interval estimation $Z=0.0131-0.0151$ for Data Set 1 by \cite{Buldgen_2024}.

Logarithmic abundance of oxygen is $8.70\pm 0.03$, carbon $8.44\pm 0.04$. These results are really close to those reported by \citeauthor{Asplund_2009} (\citeyear{Asplund_2009}, \citeyear{Asplund_2021}) and \cite{Amarsi_2021} and only marginally agree with \cite{Magg_2022}. Compared to the mixture of \cite{Asplund_2009}, we obtain a little more oxygen and a little less carbon.

Estimations of nitrogen and neon are less confident. The logarithmic abundance of nitrogen is $8.12\pm 0.08$, which is remarkably higher than all spectroscopic estimations. The neon abundance is $8.17\pm 0.09$ that agrees with accepted values by \cite{Grevesse_Noels_1993, Grevesse_Sauval_1998, Asplund_2021, Magg_2022} and exceeds \cite{Asplund_2009}. Because of neon spectroscopic estimations significantly depend on the ratio of neon to oxygen, we provide our ratio for comparison $N(Ne)/N(O)=0.29$, which appeared to be in marginal agreement with the value $0.24\pm 0.05$ by \cite{Young_2018}.

The original method of ${{\delta }_{Z}}{{\Gamma }_{1}}$ analysis, proposed by \cite{Baturin_2022}, was used for the first time to obtain information on the abundance of heavy elements in the solar convective zone. The obtained results are independent of the corresponding spectroscopic and are extremely important for solving the current question of low or high $Z$ abundances in the Sun. 

The main result of our method is the confirmation of low total mass fraction $Z$ and low content of oxygen and carbon. Another important result is the neon abundance, which is obtained in completely independent way of previous determinations. Taking into account the large uncertainty in the determination of photospheric neon from coronal measurements, our result represents a unique opportunity to refine the issue. 

Our results do not show large deviations in relation to similar determinations by other methods, but just refine them. An exception is the determination of nitrogen, which gives a significantly higher abundance than previously found ones. The accuracy estimation, taking into account methodological and model errors, turns out to be less or comparable to the accuracy of spectroscopic results of any type.


\appendix
\section{Standard Mixtures} 
\label{Appendix_StandardMixtures}  

Let us consider the standard mixtures given in the works by \cite{Grevesse_Noels_1993}, \cite{Asplund_2009}, and \cite{Magg_2022}. We call them GN93, AGSS09, MB22. SAHA-S equation of state contains eight most abundant heavy elements (C, N, O, Ne, Mg, Si, S, and Fe). Table~\ref{Table_mixtures} presents mass fractions of heavy elements for reduced mixtures when helium abundance Y=0.2485 \citep{Basu_Antia_2004} is assumed. The total mass fraction of discarded elements is about $(3-4)\cdot {{10}^{-4}}$.

%
\begin{table}
	\caption{Absolute mass fractions ${{Z}_{i}}$  for mixtures of eight elements. }
	\label{Table_mixtures}
	\begin{tabular}{lccc}     
		\hline
		Element & \multicolumn{3}{c}{$Z_i$} \\
		& GN93 & AGSS09 & MB22      \\
		\hline
		C  & 0.00310 & 0.00237 &  0.00318 \\   
		N  & 0.00095 & 0.00069 &  0.00097 \\
		O  & 0.00863 & 0.00574 &  0.00687 \\
		Ne & 0.00173 & 0.00126 &  0.00208 \\
		Mg & 0.00067 & 0.00080 &  0.00063 \\
		Si & 0.00072 & 0.00067 &  0.00080 \\
		S  & 0.00038 & 0.00031 &  0.00034 \\
		Fe & 0.00131 & 0.00129 &  0.00129 \\
		\hline
		${\left( Z/X \right)}_{\mathrm{total}}$ (original) & 0.0244  & 0.0181  & 0.0225  \\
		$Z\left( \mathrm{Mg+Si+S+Fe} \right)$                & 0.00309 & 0.00298 & 0.00305 \\
		${{Z}_{8}}=\sum\limits_{i=1}^{8}{{{Z}_{i}}}$       & 0.01750 & 0.01304 & 0.01616 \\
		${{Z}_{\mathrm{total}}}$  (original)                 & 0.0179  & 0.0134  & 0.0165  \\
		\hline
	\end{tabular}
\end{table}

\section{Methodical Errors} 
\label{Appendix_Methodical_errors}

In this Appendix, we evaluate the sensitivity of the obtained results to some of the parameters that were used in the computations. We change ${{\Gamma }_{1}}(r)$, points $\left( T(r),\rho (r) \right)$, hydrogen mass fraction, as well as the contribution of Mg, Si, S, Fe and compare the results with those given in Table~\ref{Table_Main_result}  (we note them by index 0).

\subsection{Errors of ${{\Gamma }_{1}}$ }

The inverted ${{\Gamma }_{1}}$ profile is characterized by errors $\Delta {{\Gamma }_{1}}$ about ${{10}^{-4}}$. They are shown in Fig.~\ref{Fig_G1_mod_inv} and \ref{Fig_dG1_inversion} by error bars. To evaluate how these errors might affect our ${{Z}_{i}}$ estimations, we analyze the profiles ${{\Gamma }_{1}}-\Delta {{\Gamma }_{1}}$ and${{\Gamma }_{1}}+\Delta {{\Gamma }_{1}}$. This approach gives a supremum of the error. The results are presented in Table~\ref{Table_errors_dG1}. Mass fractions vary within $\Delta {{Z}_{/}}/{{Z}_{i}}=\pm 4.5\%$ relative to their initial value (Table~\ref{Table_Main_result}), the total mass fraction of heavy elements $Z$  varies within three percent.

\begin{table}
	\caption{Mass fractions ${{Z}_{i}}$ of elements and logarithmic abundances $\log {{A}_{i}}$, obtained as a result of the least squares approximation of the inversion Z-contribution with inversion error, as well as their deviations from the main result (Table~\ref{Table_Main_result}).}
	\label{Table_errors_dG1}
	\begin{tabular}{lcccccccc}     
		\hline
		&  \multicolumn{4}{c}{${{\Gamma }_{1}}-\Delta {{\Gamma }_{1}}$} &  \multicolumn{4}{c}{${{\Gamma }_{1}}+\Delta {{\Gamma }_{1}}$} \\
		Element & $Z_i$ & $\displaystyle{\frac{\Delta Z_i}{Z_i^{(0)}}}$ & $\log A_i$ & $\Delta\log A_i$ &
		$Z_i$ & $\displaystyle{\frac{\Delta Z_i}{Z_i^{(0)}}}$ & $\log A_i$ & $\Delta\log A_i$ \\
		\hline
		C  & 0.00255 & -0.043 & 8.462 & -0.019 & 0.00234 &  0.043 & 8.425 &  0.018 \\
		N  & 0.00134 &  0.013 & 8.115 &  0.006 & 0.00137 & -0.013 & 8.127 & -0.006 \\
		O  & 0.00616 & -0.044 & 8.721 & -0.019 & 0.00564 &  0.044 & 8.683 &  0.019 \\
		Ne & 0.00222 & -0.027 & 8.177 & -0.012 & 0.00211 &  0.027 & 8.154 &  0.011 \\
		\hline
		$Z$ & \multicolumn{4}{c}{0.01525} & \multicolumn{4}{c}{0.01444} \\
		$\varepsilon $ & \multicolumn{4}{c}{$2.8\cdot {{10}^{-5}}$ } & \multicolumn{4}{c}{$4.5\cdot {{10}^{-5}}$ } \\
		\hline
	\end{tabular}
\end{table}

\subsection{The Second Inversion}

To assess possible errors, we also considered an inversion based on a different initial model \citep{Buldgen_2024}. It is calculated for the chemical mixture MB22 instead of AAG21Ne and differs from the first inversion within ${{10}^{-4}}$. Resulting abundances are presented in Table~\ref{Table_errors_secondG1}. Relative mass fraction of oxygen is increased by ${{\left( \Delta Z/Z \right)}_{\mathrm{O}}}\simeq 6\%$, carbon is practically the same, nitrogen deviation is about 16\%, and Ne 12\% by mass.

\begin{table}
	\caption{Mass fractions ${{Z}_{i}}$ of elements and logarithmic abundances $\log {{A}_{i}}$, obtained as a result of the least squares approximation of the inversion $Z$ contribution for the Inversion 2, as well as their deviations from the main result (Table~\ref{Table_Main_result}).}
	\label{Table_errors_secondG1}
	\begin{tabular}{lcccc}     
		\hline
    	Element & $Z_i$ & $\displaystyle{\frac{\Delta Z_i}{Z_i^{(0)}}}$ & $\log A_i$ & $\Delta\log A_i$ \\
		\hline
		C  & 0.00243 & -0.004 & 8.442 & -0.002 \\
		N  & 0.00113 & -0.162 & 8.044 & -0.077 \\
		O  & 0.00628 &  0.064 & 8.729 &  0.027 \\
		Ne & 0.00190 & -0.123 & 8.109 & -0.057 \\
		\hline
		$Z$ & \multicolumn{4}{c}{0.01473}  \\
		$\varepsilon $ & \multicolumn{4}{c}{$3.4\cdot {{10}^{-5}}$ }  \\
		\hline
	\end{tabular}
\end{table}

\subsection{Temperature and Density}

In the experiments described above, we used the temperature and density taken from the theoretical model described in Appendix~\ref{Appendix_Density_approximation}. To evaluate the sensitivity of our results to the choice of set $\left( T(r),\rho (r) \right)$, we performed the least squares approximation on the points $\left( T(r),\rho (r) \right)$ from theoretical models of the Sun calculated for $Z=0.018$ (high $Z$) and $Z=0.013$ (low $Z$). These models represent two extreme cases in modern solar simulations. Neon has undergone the greatest relative changes compared to Table~\ref{Table_Main_result}, ${{\left( \Delta Z/Z \right)}_{\mathrm{Ne}}}\simeq 9\%$. Nitrogen has undergone the smallest change ${{\left( \Delta Z/Z \right)}_{\mathrm{N}}}\simeq 0.8\%$. Relative changes of oxygen and carbon do not exceed $5\%$ (Table~\ref{Table_errors_TRho}). The total mass fraction of heavy elements varies within $3.5\%$.

\begin{table}
	\caption{Mass fractions ${{Z}_{i}}$ of elements and logarithmic abundances $\log {{A}_{i}}$, obtained as a result of the least squares approximation of the inversion $Z$ contribution for points $\left( T(r),\rho (r) \right)$ from high $Z$ and low $Z$ solar models, as well as their deviations from the main result (Table~\ref{Table_Main_result}).}
	\label{Table_errors_TRho}
	\begin{tabular}{lcccccccc}     
		\hline
		&  \multicolumn{4}{c}{$\left( T(r),\rho (r) \right)$ from high $Z$ model}
			&  \multicolumn{4}{c}{$\left( T(r),\rho (r) \right)$ from low $Z$ model} \\
			Element & $Z_i$ & $\displaystyle{\frac{\Delta Z_i}{Z_i^{(0)}}}$ & $\log A_i$ & $\Delta\log A_i$ &
			$Z_i$ & $\displaystyle{\frac{\Delta Z_i}{Z_i^{(0)}}}$ & $\log A_i$ & $\Delta\log A_i$ \\
			\hline
			C  & 0.00235 & -0.037 & 8.427 & -0.016 & 0.00233 & -0.047 & 8.423 & -0.021 \\
			N  & 0.00137 &  0.008 & 8.124 &  0.003 & 0.00136 &  0.006 & 8.124 &  0.003 \\
			O  & 0.00581 & -0.016 & 8.695 & -0.007 & 0.00569 & -0.037 & 8.686 & -0.016 \\
			Ne & 0.00216 & -0.003 & 8.165 & -0.001 & 0.00197 & -0.090 & 8.125 & -0.041 \\
			\hline
			$Z$ & \multicolumn{4}{c}{0.01466} & \multicolumn{4}{c}{0.01433} \\
			$\varepsilon $ & \multicolumn{4}{c}{$3.8\cdot {{10}^{-5}}$ } & \multicolumn{4}{c}{$3.9\cdot {{10}^{-5}}$ } \\
			\hline
		\end{tabular}
	\end{table}
	
\subsection{Hydrogen Mass Fraction }	

Hydrogen mass fraction is needed for computation $\Gamma _{1}^{\mathrm{HHe}}$ and basis functions $\delta _{Z}^{i}{{\Gamma }_{1}}$. In all previous computations, it is $X=0.7377$, as in the theoretical model (see Section~\ref{SbSection_ApproxRho} and Appendix~\ref{Appendix_Density_approximation}). We also carried out the computations with other $X$ values from the range of $0.72-0.74$ to evaluate the influence of this parameter on our results. The approximate range of $X$ variation was selected based on helioseismic analysis by \cite{Buldgen_2024} (see their Figure~12). The most sensitive is nitrogen ($13\%$) and neon ($21\%$). The change in oxygen is within $1\%$, carbon – $2\%$. The total mass fraction of heavy elements $Z$, as in the previous experiment, varies within 3.5 percent (Table~\ref{Table_errors_X}).

\begin{table}
	\caption{Mass fractions ${{Z}_{i}}$ of elements and logarithmic abundances $\log {{A}_{i}}$, obtained as a result of the least squares approximation of the inversion $Z$ contribution with inversion error, as well as their deviations from the main result (Table~\ref{Table_Main_result}).}
	\label{Table_errors_X}
	\begin{tabular}{lcccccccc}     
		\hline
		&  \multicolumn{4}{c}{$X=0.72$} &  \multicolumn{4}{c}{$X=0.74$} \\
		Element & $Z_i$ & $\displaystyle{\frac{\Delta Z_i}{Z_i^{(0)}}}$ & $\log A_i$ & $\Delta\log A_i$ &
		$Z_i$ & $\displaystyle{\frac{\Delta Z_i}{Z_i^{(0)}}}$ & $\log A_i$ & $\Delta\log A_i$ \\
		\hline
		C  & 0.00249 & 0.019 & 8.463 & 0.019 & 0.00244 & -0.002 & 8.442 & -0.002 \\
		N  & 0.00118 & -0.130 & 8.071 & -0.050 & 0.00138 & 0.017 & 8.127 & 0.006 \\
		O  & 0.00597 & 0.011 & 8.718 & 0.015 & 0.00589 & -0.001 & 8.701 & -0.002 \\
		Ne & 0.00172 & -0.207 & 8.076 & -0.090 & 0.00222 & 0.027 & 8.176 & 0.010 \\
		\hline
		$Z$ & \multicolumn{4}{c}{0.01433} & \multicolumn{4}{c}{0.01491} \\
		$\varepsilon $ & \multicolumn{4}{c}{$2.9\cdot {{10}^{-5}}$ } & \multicolumn{4}{c}{$3.7\cdot {{10}^{-5}}$ } \\
		\hline
	\end{tabular}
\end{table}

\subsection{Contribution of Mg, Si, S, and Fe}

In this experiment, we changed the contribution of Mg, Si, S, and Fe by 10 percent. The scale of the change is taken from the error in determining iron, the most abundant element in this group, according to data given in \cite{Grevesse_Noels_1993, Asplund_2009, Magg_2022} . Oxygen practically does not feel this change ($0.4\%$). Carbon deviates by $8\%$, nitrogen by $1\%$, and neon undergoes the greatest relative changes - by $11\%$. The total mass fraction $Z$  changes by 3 percent (Table~\ref{Table_errors_MgSiSFe}).

\begin{table}
\caption{Mass fractions ${{Z}_{i}}$ of elements and logarithmic abundances $\log {{A}_{i}}$, obtained as a result of the least squares approximation of the inversion $Z$ contribution with changed contribution of Mg, Si, S, and Fe by 10 percent, as well as their deviations from the main result (Table~\ref{Table_Main_result}).}
\label{Table_errors_MgSiSFe}
\begin{tabular}{lcccccccc}     
\hline
&  \multicolumn{4}{c}{$\Delta Z(Mg,Si,S,Fe)=+10\%$} & \multicolumn{4}{c}{$\Delta Z(Mg,Si,S,Fe)=-10\%$} \\
Element & $Z_i$ & $\displaystyle{\frac{\Delta Z_i}{Z_i^{(0)}}}$ & $\log A_i$ & $\Delta\log A_i$ &
$Z_i$ & $\displaystyle{\frac{\Delta Z_i}{Z_i^{(0)}}}$ & $\log A_i$ & $\Delta\log A_i$ \\
\hline
C  & 0.00226 & -0.077 & 8.409 & -0.035 & 0.00263 &  0.077 & 8.476 &  0.032 \\
N  & 0.00137 &  0.012 & 8.126 &  0.005 & 0.00134 & -0.012 & 8.116 & -0.005 \\
O  & 0.00593 &  0.004 & 8.704 &  0.002 & 0.00588 & -0.004 & 8.701 & -0.002 \\
Ne & 0.00192 & -0.115 & 8.113 & -0.053 & 0.00241 &  0.115 & 8.213 &  0.047 \\
\hline
$Z$ & \multicolumn{4}{c}{0.01445} & \multicolumn{4}{c}{0.01524} \\
$\varepsilon $ & \multicolumn{4}{c}{$3.3\cdot {{10}^{-5}}$} & \multicolumn{4}{c}{$4.0\cdot {{10}^{-5}}$ } \\
\hline
\end{tabular}
\end{table}

\subsection{Conclusion on Error Estimations}

\begin{figure} 
\centerline{\includegraphics[width=1.0\textwidth,clip=]{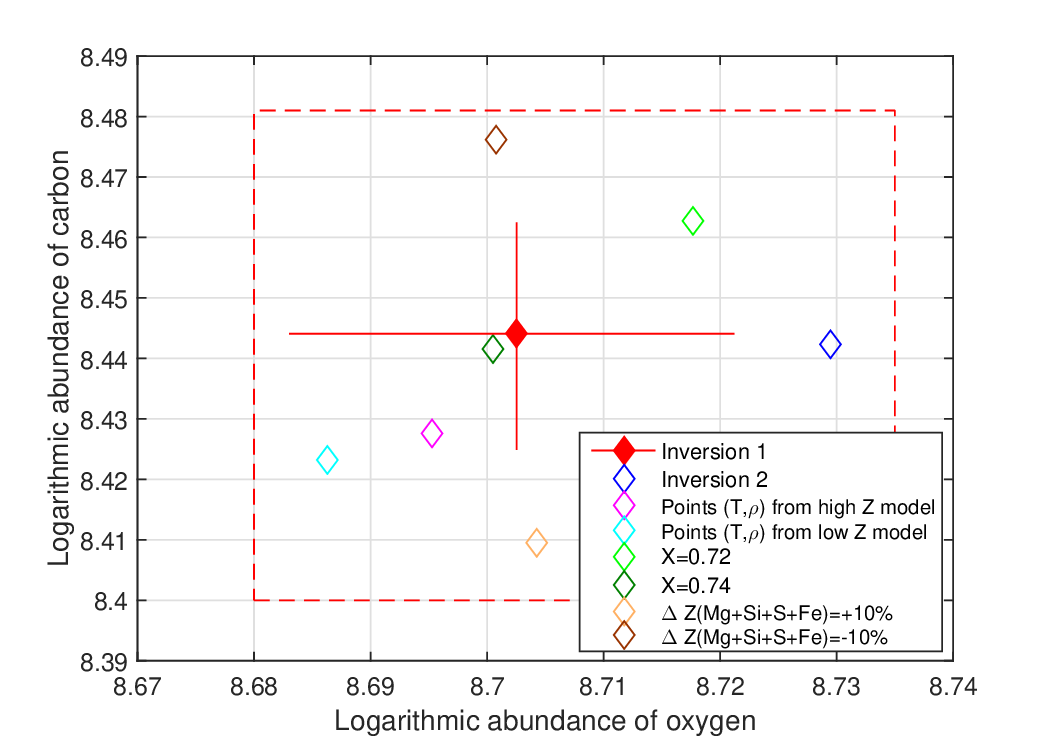}}
\caption{Logarithmic abundances of oxygen and carbon at different input parameters (${{\Gamma }_{1}}$, points $\left( T,\rho  \right)$, hydrogen mass fraction $X$  in the plasma, and the contribution of Mg, Si, S, and Fe).}
\label{Fig_lgA_CO_errors}
\end{figure}

\begin{figure} 
	\centerline{\includegraphics[width=1.0\textwidth,clip=]{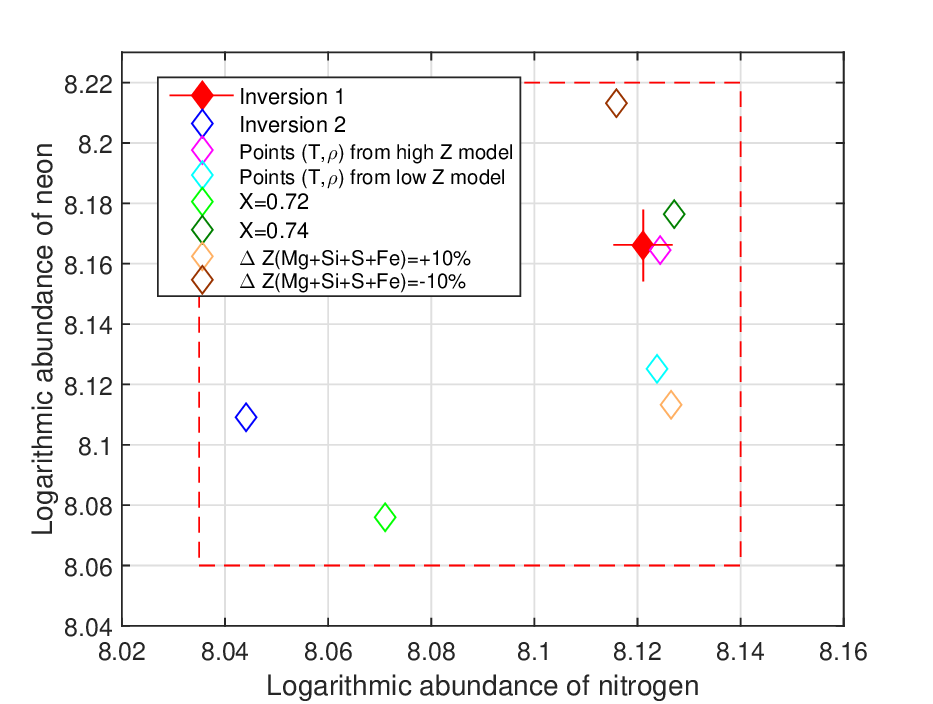}}
	\caption{Logarithmic abundances of nitrogen and neon at different input parameters (${{\Gamma }_{1}}$, points $\left( T,\rho  \right)$, hydrogen mass fraction $X$  in the plasma, and the contribution of Mg, Si, S, and Fe).}
	\label{Fig_lgA_NNe_errors}
\end{figure}

All estimations described in Appendix~\ref{Appendix_Methodical_errors} are shown in Figures~\ref{Fig_lgA_CO_errors} and \ref{Fig_lgA_NNe_errors} by colored empty diamonds. The red filled diamond is the main result (see Table~\ref{Table_Main_result}). Red dashed rectangle, as in Figures~\ref{Fig_lgA_O_C} and \ref{Fig_lgA_N_Ne}, outlines the region where our experiments fall into. The red cross shows the estimations due to errors $\Delta {{\Gamma }_{1}}$ obtained in the inversion procedure. The spread of estimations of oxygen $\Delta \log {{A}_{\mathrm{O}}}$ is about 0.05. The greatest effect is caused by the profile ${{\Gamma }_{1}}\left( r \right)$ and points $\left( T(r),\rho (r) \right)$. Carbon spread is 0.08. The maximum deviations are caused by the uncertainty of the total contribution of Mg, Si, S, and Fe.

The logarithmic abundance of nitrogen is most sensitive to the choice of inversion profile ${{\Gamma }_{1}}\left( r \right)$. Although the profiles differ very little, within $10^{-4}$, nitrogen reacts to these minor changes. The maximum spread in nitrogen is about 0.09. Neon is most sensitive to the hydrogen mass fraction $X$, the spread of results is 0.10.

We consider the maximum deviation of experimental points from our central result as an estimation of error. For oxygen it turns out $8.70\pm 0.03$, for carbon $8.44\pm 0.04$, for nitrogen $8.12\pm 0.08$, for neon $8.17\pm 0.09$.

\section{Accuracy of Approximation of Inverted Density Profile } 
\label{Appendix_Density_approximation} 

\begin{figure} 
\centerline{\includegraphics[width=1.0\textwidth,clip=]{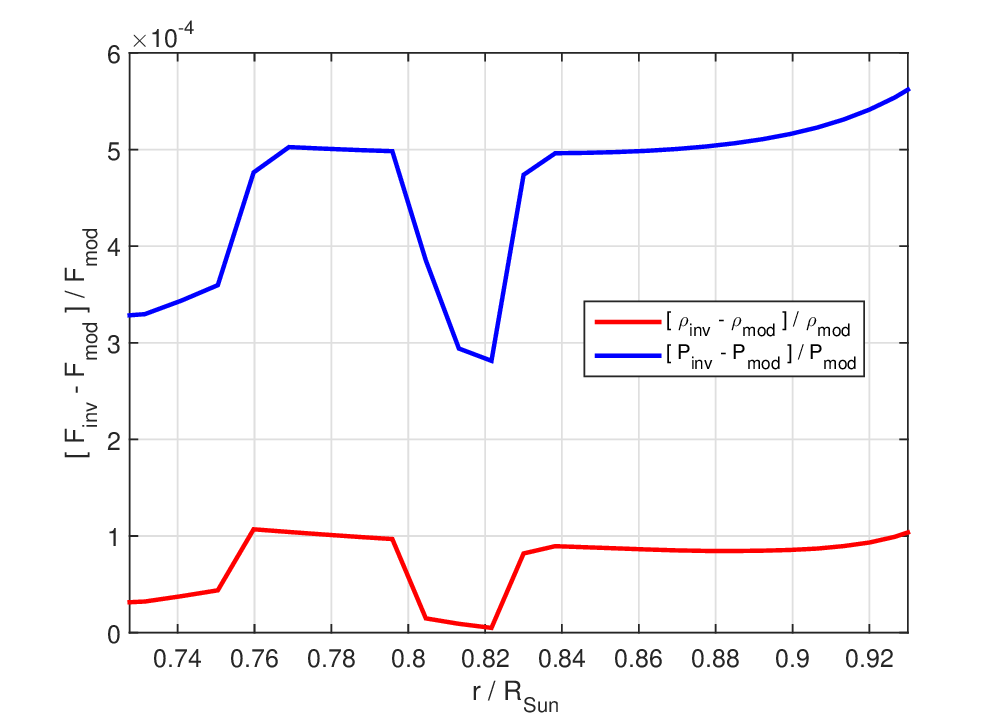}}
\caption{Deviation of theoretical density and pressure from inverted ones.}
\label{Fig_dRho_dP_inv_mod}
\end{figure}

The inversion procedure gives profiles of ${{\Gamma }_{1}}$, density and pressure. However, to apply our method, it is also necessary to know the temperature and mass fraction of hydrogen. In addition, the inverted density and pressure profiles contain errors of unknown nature. Therefore, we selected a theoretical model in which the density profile is smooth and coincides with the inverted one with a certain accuracy. The model that we use has the following parameters: $m_{75}=0.98236$, $X=0.7377$, $Y=0.2479$, $Z=0.0144$. Figure~\ref{Fig_dRho_dP_inv_mod} shows the deviation of the theoretical density and pressure from the inversion ones. The theoretical density agrees with the inversion with an accuracy of ${{10}^{-4}}$ (red curve), and pressure – within $6\cdot {{10}^{-4}}$ (blue curve).  We consider the difference between the approximation model and the inversion to be small in view of estimates of heavy elements content. Additional estimates show that they are inside the error square in Figures~\ref{Fig_lgA_CO_errors} and \ref{Fig_lgA_NNe_errors}.
In the general case, such a model is not the only one. The sensitivity of our method to the given $T(r)$, $\rho (r)$, and X is examined in Appendix~\ref{Appendix_Methodical_errors}.

%



\begin{authorcontribution}
V.A.B. and A.V.O. proposed the idea and performed computation on the analysis of Z contribution to adiabatic exponent. A.S.V. was responsible for solar modeling computations. V.K.G. and I.L.I. provide computations for SAHA-S EOS data. G.B., A.N., and R.S. were responsible for inversion modeling data, integration of the SAHA-S equation of state in the Liège stellar evolution code, CLES and oscillation computations for the solar calibrated models. All the co-authors contributed to the interpretation of the results.
\end{authorcontribution}
\begin{fundinginformation}
G. Buldgen acknowledges fundings from the Fonds National de la Recherche Scientifique (FNRS) as a postdoctoral researcher. The study by V.K. Gryaznov is conducted under the government contract for fundamental research registration number 124020600049-8.
\end{fundinginformation}
\begin{dataavailability}
No datasets were generated or analysed during the current study.
\end{dataavailability}
\begin{ethics}
\begin{conflict}
The authors declare no competing interests.
\end{conflict}
\end{ethics}

%
%
\bibliographystyle{spr-mp-sola}
\bibliography{bibliography}  
%
%
%
%

\end{document}